\documentclass[conference]{IEEEtran}
\IEEEoverridecommandlockouts

\usepackage{cite}
\usepackage{amsmath,amssymb,amsfonts}
\usepackage{subcaption}
\usepackage{tikz}
\usetikzlibrary{positioning,decorations.pathmorphing,fit,calc}

\usepackage{pgfplots}
\usepgfplotslibrary{units}

\usepackage{siunitx}
\usepackage{algorithmic}
\usepackage{graphicx}
\usepackage{textcomp}
\usepackage{xcolor}
\usepackage{hyperref}
\usepackage{orcidlink}
\def\BibTeX{{\rm B\kern-.05em{\sc i\kern-.025em b}\kern-.08em
T\kern-.1667em\lower.7ex\hbox{E}\kern-.125emX}}
\usepackage{mathtools}

\newcommand{\defeq}{\stackrel{\mathclap{\mbox{\scriptsize def}}}{=}}

\begin{document}

\title{Predictable Modelling and Analysis of Software-defined Vehicle Implementations\\
  \thanks{European Chips Joint Undertaking under Framework Partnership Agreement No 101139789 (HAL4SDV)}
}

\author{
  \IEEEauthorblockN{
    Pavlo Tokariev\textsuperscript{1} \orcidlink{0000-0002-4603-4770}, 
    Yosri Ayari\textsuperscript{1} \orcidlink{0009-0004-3135-9101}, 
    and Julien Deantoni\textsuperscript{1,2} \orcidlink{0000-0001-6962-7846}
  }
  \IEEEauthorblockA{
    \textsuperscript{1}\textit{Inria, Kairos Team}, Sophia-Antipolis, France \\
    \textsuperscript{2}\textit{Université Côte d'Azur, I3S, CNRS}, Nice, France \\
    Email: \{pavlo.tokariev, abdelkader-yosri.ayari\}@inria.fr, julien.deantoni@univ-cotedazur.fr
  }
}

\maketitle

\begin{abstract}
Software-Defined Vehicles (SDVs) rely on middleware-based communication and hardware abstraction mechanisms that introduce temporal uncertainty affecting end-to-end timing guarantees. Previous work proposed probabilistic architectural models for early timing analysis, but the representativeness of these abstractions with respect to SDV implementations remained unclear.
This paper presents an experimental framework combining probabilistic design-time timing analysis with a monitored Kuksa-based implementation. The same reaction-time analysis is applied both to simulation and implementation traces, enabling direct comparison between predicted and observed timing behaviour. We additionally introduce a comparison methodology separating conservative coverage from predictive fidelity of timing distributions.
The results show that the proposed abstractions remain representative under different middleware load conditions while preserving conservative timing guarantees, supporting incremental timing verification approaches for SDV platforms.
\end{abstract}


\section{Introduction}

Software-Defined Vehicles (SDVs) aim to improve the agility and scalability of automotive software development through a strong decoupling between software and hardware enabled by a Hardware Abstraction Layer (HAL). While this abstraction improves portability and modularity, it also introduces temporal uncertainty due to indirect hardware access and middleware-based communication. Managing this uncertainty during early design stages is therefore critical to preserve end-to-end timing guarantees throughout system deployment and evolution~\cite{jezequel:hal-04064771}.

Previous work~\cite{ICTERI2025} proposed methods for early timing analysis of functional chains based on high-level architectural models. Systems are represented as interacting software components communicating through signals and accessing hardware-related data exposed through the HAL using standards such as the Vehicle Signal Specification (VSS). Components are annotated with probabilistic timing specifications capturing deployment-induced uncertainty, enabling analysis of end-to-end reaction-time distributions during design.

\begin{figure}[t]
  \centering
  \begin{tikzpicture}[node font=\tiny]
    \tikzset{box/.style={draw,rectangle,minimum width=1cm,minimum height=0.5cm,text width=1.2cm,text centered}}
    \node[box] (model) at (-3,0) {SDV Model};
    \node[box] (impl) at (-3,-2) {Implementation};
    \node[box] (simtraces) at (0,0) {Traces};
    \node[box] (param) at (0,-1) {Timing uncertainty};
    \node[box] (impltraces) at (0,-2) {Traces};
    \node[box] (res1) at (3,0) {Reaction time distribution};
    \node[box] (res2) at (3,-2) {Reaction time distribution};

    \begin{scope}[shorten <= 2pt, shorten >= 2pt]
      \draw[->] (model) -- node[above] {Simulate} (simtraces) ;
      \draw[->] (impl) -- node[above] {Monitor} (impltraces) ;
      \draw[->] (simtraces) -- node[above] {Analysis} (res1);
      \draw[->] (impltraces) -- node[above] {Analysis} (res2);
      \draw[->] (impltraces) -- node[right] {Extract} (param);
      \draw[->,dotted] (impl) -- node[left] {Implements} (model);
      \draw[decorate, decoration=snake,red] (res1) -- node[right,text width = 2cm,xshift=0.1cm,text width=1cm] {Fidelity assessment} (res2);
      \draw[->,red] (res1) edge[out=120,in=60,looseness=0.5] node[above] {Feedback} (model);
      \draw[->,red] (param) edge[out=180,in=-60] node[right,node font=\tiny,text width=1cm,xshift=0.1cm,yshift=0.1cm] {Inject} (model);
    \end{scope}
  \end{tikzpicture}
  \caption{Overview of the approach}
  \label{fig:overview}
\end{figure}
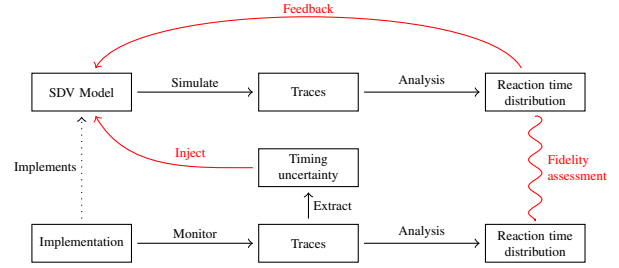

However, an important open question remains regarding the representativeness of these abstractions when confronted with concrete implementations deployed on industrial SDV platforms such as Eclipse Kuksa~\cite{KUKSA}. To investigate this question, this work proposes an experimental framework combining a Kuksa-based implementation with probabilistic design-time timing analysis (Fig.~\ref{fig:overview}). The framework includes configurable software components, mocked sensors and actuators, and monitoring capabilities allowing reconstruction of functional-chain reaction times from observed executions. The same analysis pipeline is then applied both to simulated traces generated from the architectural model and to monitored implementation traces, enabling direct comparison between predicted and observed timing behaviour.

To assess the validity of the proposed abstractions, this paper introduces a comparison methodology separating conservative coverage from predictive fidelity of the timing distributions. Rather than requiring exact statistical equivalence, the approach evaluates whether the design-time model safely contains observed executions while remaining representative of practical implementation behaviour.

The results show that probabilistic architectural timing analysis remains representative of implementation behaviour under different middleware load conditions while preserving conservative timing guarantees. More importantly, the experiments indicate that middleware-induced timing uncertainty exhibits structured dependencies on communication activity, particularly on the number of publishers and subscribers interacting with the databroker. These observations suggest that timing verification may be performed incrementally by re-analysing applications individually under updated middleware uncertainty conditions rather than requiring complete system-wide reanalysis.

\section{Related Work}
\label{sec:rel-work}

End-to-end timing analysis of Robot Operating System 2 (ROS2) systems, based on the Data Distribution Service (DDS), has been widely studied for distributed real-time applications, similar to the ones of SDVs.
Casini et al.~\cite{casini2019response} proposed one of the first formal response-time analyses for ROS2 processing chains under reservation-based scheduling.
More recent work extended these analyses to single~\cite{teper2022} and multi-executor systems~\cite{teper2024end}, timing jitter management~\cite{abaza2024managing}, and DDS-based communication effects~\cite{sciangula2025end}.
These approaches mainly focus on deterministic or worst-case timing guarantees.
Probabilistic timing analysis has also been explored to capture execution-time variability and deployment uncertainty in complex real-time systems.
Recent work proposed probabilistic architectural models based on Modular Real-Time Clock Constraint Specification Language (MRTCCSL)~\cite{tokariev_2024dec} for early SDV functional-chain timing analysis under uncertainty~\cite{ICTERI2025}.
In contrast to existing work, this paper focuses on the representativeness of probabilistic architectural timing abstractions with respect to monitored SDV implementations.
Rather than only deriving analytical bounds, we compare simulated and implementation-derived reaction-time distributions using the same analysis pipeline while explicitly separating conservative coverage from predictive fidelity.

\section{System Model and Probabilistic Timing Specification}
\label{sec:model}

We model the behaviour of an SDV system in two parts.
The first part describes the HAL, an abstraction of the hardware platform, as a set of hardware-related signals available to the applications, allowing either reading or writing (i.e. sensors and actuators, respectively).
Signals specify their triggering policies (periodic or event-triggered) as well as latency, defining the duration between the activation and the availability of the corresponding data to the application for a sensor, or between the activation and the action for an actuator.

The second part describes the application under study and consists of interconnected software components implementing the functional features of the vehicle.
These components realize the data flow to and from the HAL and generally define a control loop.
A component consists of a set of input and output signals (either connected to the HAL or local to the application), a periodic or event-triggered activation policy, and an execution time.
Each input signal in the component is annotated with its consumption pattern, either as a variable (multiple reads between writes return the same value) or as a queue (each read consumes a value from the queue).

All timing parameters are expressed as independent normal distributions, optionally bounded to enforce physically meaningful timing values.
The mean represents the nominal timing specification, while the standard deviation represents the timing uncertainty introduced by the deployment platform.
This uncertainty may be calibrated from observed executions or estimated from engineering expertise.

To express system safety requirements directly within the design, we define timing constraints on functional chains representing the system reaction time~\cite{Feiertag2008ACF}.
A functional chain specification consists of a sequence of signals and application components.
Typically, a functional chain starts with a sensor and ends with an actuator.
The chain identifies the specific path followed by the propagating data and defines the corresponding reaction time between component activations, which is later analysed as a probability distribution and checked against the timing requirements.

For design-time analysis, the timing parameters and triggering policies of the application components and HAL elements are encoded as MRTCCSL specifications, capturing the system timing behaviour.
In this description, several abstractions are assumed, such as synchronous communication between application components (thereby neglecting communication delays), known component activation offsets, and the underlying scheduling algorithm.

MRTCCSL~\cite{tokariev_2024dec} is a constraint-based language over logical and real-time clocks.
It supports both stochastic timing expressions and quantitative temporal relations between events.
These features allow flexible modelling of timing behaviour including uncertainty, while efficiently generating representative execution traces of the SDV system.

In this work, we use an example of a system implementing cruise control functionality.
The cruise control maintains a constant vehicle speed during its operation, typically above a predefined speed threshold.
The target speed is implicitly set by the user by accelerating the vehicle to the desired speed and enabling the system.
The functionality is disabled whenever the brake is applied.

Our model and implementation of the system consists of 4 sensors, 2 actuators, and 9 application components.
Sensors are periodic, actuators are event-triggered, and components are mixed: components responsible for reading and writing to HAL signals are event-triggered, while others are periodic.
The complete description of the system, its implementation and experimental results are available in 
\url{https://github.com/jdeantoni/SDV_experiments_VPPC_26}

\section{System Implementation and Monitoring}
\label{sec:impl}

To evaluate whether the proposed SDV and HAL abstractions are representative of a real deployment, we implement an execution framework composed of software components, including simulated sensors and actuators on the hardware side, and reader, controller, and writer components on the application side.

The framework follows the SDV architecture by decoupling applications from the hardware platform through a HAL, whose current implementation relies on the Kuksa databroker for communication.

Software components are parameterized using the same attributes as in the model, namely triggering policies, execution times, and input/output signals. Periodic and event-triggered activations are supported, while communication between components is performed through Kuksa subscriptions and publications. In contrast to the probabilistic analysis model, the implementation executes concrete timing configurations and does not inject the uncertainty distributions specified at design time. This enables execution of systems directly derived from the model while preserving the activation and communication semantics of the specification.

The implementation framework records execution traces of component activations and databroker interactions. From these traces, we reconstruct causal relations between produced and consumed data and extract observed timing behaviours, including execution times and component-to-component communication latencies along the functional chain. Aggregating traces from multiple runs allows estimating the timing uncertainty introduced by the middleware, operating system, and deployment platform. These measured timing distributions can then be compared with the assumptions made at design time or injected back into the system model to better represent the uncertainty of the deployment platform.

\section{Reaction Time Analysis}
\label{sec:sim-analysis}

We implement a probabilistic reaction time analysis of the proposed system model in two steps: execution trace generation and functional chain identification. This separation allows to analyze traces originating either from a simulation or external sources, such as implementation monitoring.

Using the functional chain specification and communication semantics between system elements, we identify functional chain instances within the generated traces.
The communication semantics establishes the causal relationships between the start and completion of a chain and enables computation of the corresponding reaction times. The detailed functional-chain reconstruction procedure is omitted for brevity.

Finally, reaction times are collected from multiple traces and aggregated into a probabilistic distribution represented as a histogram.
This workflow enables a unified probabilistic analysis of both simulated and implementation-derived executions.
Fig.~\ref{fig:sim-result} and Fig.~\ref{fig:impl-result} present the resulting reaction time distributions for the cruise control example simulation and implementation respectively.

\section{Probabilistic Timing Analysis Assessment}
\label{sec:conceptual-positioning}

Classical statistical metrics such as Jensen-Shannon divergence, Total Variation distance, Bhattacharyya distance, or Wasserstein distance evaluate global statistical similarity between distributions~\cite{gibbs2002choosing}.
These metrics are widely used in statistical model validation, uncertainty quantification, and probabilistic verification.
However, a design-time (timing) specification is not intended to exactly represent implementation behaviour; instead, it should safely over-approximate the set of admissible executions while remaining predictive of (future) observed timing properties.
Consequently, purely symmetric distribution distances may incorrectly penalize conservative specifications because they treat additional admissible probability mass as a statistical error.

For this reason, the proposed assessment framework separates two complementary dimensions:
\begin{enumerate}
\item \textbf{Coverage}, evaluating whether implementation behaviour remains safely contained within the specification.
\item \textbf{Fidelity}, evaluating whether the specification remains predictive of observed implementation timing behaviours.
\end{enumerate}

\subsection{Coverage Property}

The coverage analysis evaluates whether implementation observations remain contained within the specification bounds, and thus whether the timing requirement validation performed at design time remains valid during operation.
We define the coverage property as the subset relation between the distribution supports $\mathrm{supp}(D_i)\subseteq_{\epsilon_c} \mathrm{supp}(D_s) \defeq \left|\mathrm{supp}(D_i) \setminus \mathrm{supp}(D_s) \right| \leq \epsilon_c$, where a support is a realizable subset of the distribution domain $\mathrm{supp}(D) \defeq \{t \mid D(t)>0\}$.

In practice, empirical supports are highly sensitive to finite sampling, discretization, and rare events.
The assessment framework therefore evaluates coverage up to a probability mass $\epsilon_c$ considered as negligible if uncovered. When $\epsilon_c = 0$, the specification is strictly conservative.
For $\epsilon_c > 0$, the framework tolerates a negligible uncovered implementation mass while preserving the interpretation of the specification as a conservative abstraction.

\subsection{Tolerance-Aware Fidelity}

Implementation traces and design-time timing analyses may differ due to measurement jitter, synchronization offsets, calibration errors, discretization artifacts, or bounded scheduling variability.

To account for these effects, the assessment framework uses the continuous moving average with a tolerance parameter $\epsilon$:
\begin{equation*}
\widetilde D^{\,\epsilon}(v) = \frac{1}{2\epsilon}\int^{v+\epsilon}_{v-\epsilon} D(t)\,dt
\end{equation*}

This operation is related to kernel smoothing and robust statistical comparison techniques~\cite{silverman1986density}.
The resulting divergence-versus-tolerance curve of the smoothed simulation and implementation reaction time distributions provides the additional explainability:
\begin{itemize}
\item A rapid divergence decrease suggests mainly local timing offsets.
\item A persistent divergence plateau suggests structural behavioural mismatch.
\end{itemize}

The tolerance-aware metrics are therefore not intended as strict information-theoretic divergences, but rather as operational robustness indicators for practical timing validation.

\subsection{Implementation-Conditioned Fidelity}

Classical symmetric fidelity measures (under tolerance or not) evaluate the global similarity between specification and implementation distributions.
Such measures are particularly informative when the collected implementation traces are considered exactly representative of the operational timing behaviour.

However, a conservative specification may intentionally contain admissible executions that are not observed in a particular implementation, deployment configuration, or execution campaign.
To avoid treating these admissible yet unobserved behaviours as prediction errors, the framework additionally evaluates fidelity restricted to regions where implementation behaviour is effectively observed.
This implementation-conditioned fidelity is especially useful when the observed executions only exercise part of the admissible timing behaviour described by the specification.

The resulting normalized distributions correspond to conditional distributions restricted to observed implementation regions.
The statistical metrics obtained from these distributions should therefore be interpreted as the \emph{quality of prediction considering observed implementation executions} rather than as a global probabilistic similarity measure.
Also, to preserve visibility of specification conservatism, the framework additionally reports 
$M_{\mathrm{extra\_spec}} = \int_{t\notin \mathrm{supp}(D_i)} D_s(t)\,dt$, which quantifies specification probability mass not covered by implementation observations.

This ``implementation-conditioned'' comparison can also be combined with the tolerance-aware analysis introduced previously.

\subsection{Practical Interpretation}

The proposed assessment framework ultimately evaluates whether the specification is simultaneously:
\begin{enumerate}
\item Sufficiently conservative to safely cover implementation behaviour.
\item Sufficiently predictive of observed execution timing properties (either implementation conditioned or not).
\end{enumerate}

The fidelity criteria may additionally be analysed with respect to tolerance in order to distinguish local timing discrepancies from structural behavioural differences.

A specification may therefore be:
\begin{itemize}
\item Statistically imperfect yet operationally safe.
\item Highly conservative yet weakly predictive.
\item Locally accurate but globally non-conservative.
\end{itemize}

The proposed decomposition makes these situations explicitly distinguishable instead of collapsing them into a single similarity score.



\section{Evaluation}
\label{sec:evaluation}

We evaluate the proposed probabilistic design-time timing analysis on the cruise control implementation introduced previously in Section~\ref{sec:model}.
To assess whether the analysis provides a sound and representative abstraction of implementation behaviour, we use the distribution comparison methodology introduced in Section~\ref{sec:conceptual-positioning}.

Two deployment conditions are considered.
The first one corresponds to a nominal execution where the Kuksa databroker is only used by the cruise control application.
The second one introduces an additional synthetic load composed of multiple publishers and subscribers interacting concurrently with the databroker.
For both configurations, execution traces are collected from the implementation, timing uncertainties introduced by the platform are extracted and injected back into the design-time model, and the resulting probabilistic functional-chain reaction-time distributions are compared with the observed implementation behaviour.

For the nominal deployment, the probabilistic timing analysis remains globally representative of the implementation behaviour while preserving conservative timing guarantees.
The symmetric comparison yields a Jensen-Shannon divergence of $0.163$, indicating a moderate global statistical discrepancy between the distributions.
At the same time, the Bhattacharyya coefficient reaches $0.862$, showing a strong overlap between specification and implementation timing behaviours.
From the coverage perspective, the implementation support is almost entirely contained within the specification support, with only $0.0031$ uncovered implementation probability mass.
This indicates that the design-time timing analysis remains conservative with respect to the observed executions.
The tolerance-aware analysis further suggests that a substantial part of the discrepancy originates from local timing variability rather than structural behavioural mismatch.
Indeed, the Jensen-Shannon divergence decreases from $0.163$ to $0.097$ when introducing a tolerance of $\qty{0.4}{ms}$ (approximately $1\%$ of the mean reaction time).
The implementation-conditioned analysis additionally highlights the conservative nature of the specification.
Approximately $14.6\%$ of the specification probability mass corresponds to admissible but unobserved in implementation behaviour.
After introducing tolerance, the restricted Jensen-Shannon divergence decreases to approximately $0.040$, indicating strong predictive fidelity within the observed operational region.

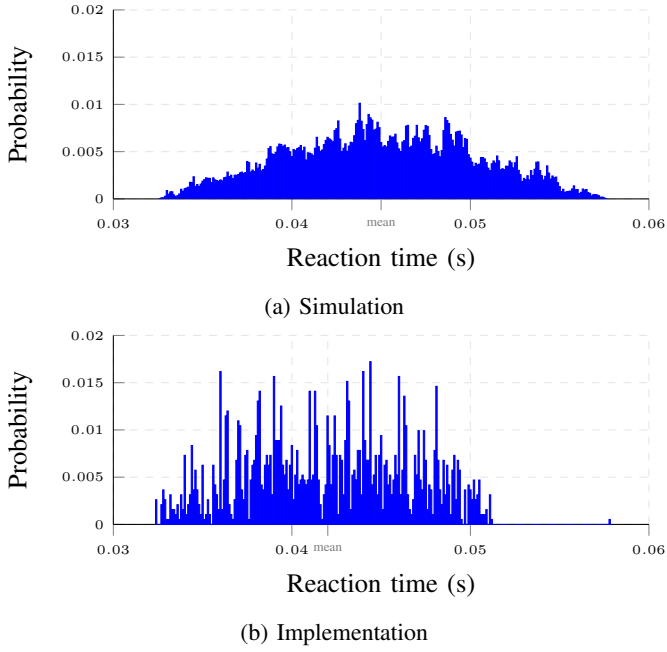
\begin{figure}[ht]
    \centering
    \pgfplotsset{
        shared_style/.style={
            width=0.8\linewidth,
            height=2.5cm,
            scale only axis, 
            ybar,
            bar width=0.5pt, 
            ymin=0, ymax=0.02,
            xmin=0.03, xmax=0.06,
            xtick={0.03, 0.04, 0.05, 0.06},
            ytick={0, 0.005, 0.01, 0.015, 0.02},
            scaled ticks=false,
            xticklabel style={
                /pgf/number format/fixed,
                /pgf/number format/precision=3,
                font=\tiny
            },
            yticklabel style={
                /pgf/number format/fixed,
                /pgf/number format/precision=3,
                font=\tiny
            },
            ylabel={Probability},
            xlabel={Reaction time (s)},
            axis lines*=left,
            grid=major,
            grid style={dashed, gray!20}
        }
    }
    \begin{subfigure}{\linewidth}
        \begin{tikzpicture}
          \begin{axis}[
              shared_style, 
              extra x ticks={0.044996}, 
              extra x tick labels={mean}, 
              extra x tick style={grid=major, tick label style={anchor=north, font=\tiny, color=gray}}
            ]
            \addplot+[blue, fill=blue] table [x=bin, y=total, col sep=comma] {./sim.csv};
          \end{axis}
        \end{tikzpicture}
        \caption{Simulation}
        \label{fig:sim-result}
    \end{subfigure}
    \begin{subfigure}{\linewidth}
        \begin{tikzpicture}
          \begin{axis}[
              shared_style, 
              extra x ticks={0.042012}, 
              extra x tick labels={mean}, 
              extra x tick style={grid=major, tick label style={anchor=north, font=\tiny, color=gray}}
            ]
            \addplot+[blue, fill=blue] table [x=bin, y=total, col sep=comma] {./impl.csv};
          \end{axis}
        \end{tikzpicture}
        \caption{Implementation}
        \label{fig:impl-result}
    \end{subfigure}

    \caption{Reaction time distributions from trace analysis}
    \label{fig:reaction-comparison}
\end{figure}

Fig.~\ref{fig:reaction-comparison} illustrates the reaction-time distributions obtained from simulation and implementation under nominal execution conditions.
Although the implementation histogram appears visually more irregular and sparse due to finite execution sampling and runtime variability, both distributions exhibit similar support regions and global timing structure.
This motivates the need for probabilistic and tolerance-aware comparison metrics rather than relying solely on direct visual inspection of histograms.

Under synthetic middleware load, the probabilistic design-time timing analysis similarly remains representative of the observed implementation behaviour while preserving conservative coverage.
The global Jensen-Shannon divergence reaches $0.119$, indicating a limited global statistical discrepancy, while the Bhattacharyya coefficient reaches $0.899$, again indicating strong overlap between specification and implementation distributions.
The specification fully contains the observed implementation support, with no uncovered implementation probability mass.
The implementation-conditioned comparison yields a restricted Jensen-Shannon divergence of $0.066$, showing that the probabilistic timing analysis remains predictive within the observed operational region even under additional middleware activity.
The tolerance-aware analysis indicates that most discrepancies correspond to local temporal shifts rather than structural behavioural differences.
The Jensen-Shannon divergence decreases to approximately $0.1$ when introducing a tolerance of $\qty{0.4}{ms}$.

Overall, the results show that the proposed probabilistic timing analysis preserves conservative timing guarantees while remaining representative of observed implementation behaviour under different deployment conditions.
The proposed comparison methodology additionally helps distinguish local runtime variability from structural mismatches between specification and implementation timing behaviours.

\section{Discussion}
\label{sec:discussion}

While the results demonstrate strong alignment between the architectural model and the implementation, certain factors could influence the generality and precision of our results.
A potential concern is the use of independent normal distributions for timing specification, which may not fit all observable middleware jitter or execution patterns.
However, the underlying MRTCCSL simulation engine is distribution-agnostic; the methodology can be directly extended to other stochastic profiles (e.g., empirical or heavy-tailed distributions) without structural changes.
Furthermore, our model enforces physical bounds to ensure plausible timing values (e.g., non-negative latencies).
In contrast, the distribution independence may lead to the loss in the design-time analysis precision in the case if the independence assumption does not hold by the implementation.
Which is however generally acceptable as the ``over''-exploration of timing behaviour by the simulation guarantees that the analysis is conservative and thus the reaction time requirements remain sound when demonstrated at design-time.

Regarding instrumentation, the monitoring overhead is negligible, with observed reaction time means remaining within $1\%$ of the simulation, ensuring that the ``probe effect'' did not bias our fidelity assessment.

The experimental results indicate that, for the considered SDV implementation and middleware configuration, the proposed probabilistic timing model remains representative of observed execution behaviour under varying load conditions, while preserving conservative coverage of implementation traces. However, these findings are obtained within a controlled experimental setting and do not yet establish general applicability across arbitrary SDV platforms or workload regimes. In particular, while a dependency between middleware load and timing uncertainty is consistently observed, this relationship remains empirical rather than formally characterized.

Within this scope, the results suggest that deployment-induced timing variability is not purely unstructured, but exhibits regular dependencies on observable system properties such as communication load. This observation opens the possibility of approximating platform effects through surrogate models that map system configuration to timing uncertainty parameters.
This issue is particularly relevant in SDV architectures, where applications interact through a shared communication middleware such as Kuksa databroker, making their timing behaviour dependent on global platform activity rather than only local execution logic.

The results of our probabilistic timing analysis suggest a way to relax this dependency.
When deployment-induced uncertainty is properly captured, design-time analysis remains representative of implementation behaviour.
This indicates that when the effect of the platform can be abstracted through uncertainty parameters, full-system reanalysis for each deployment can be avoided.
Based on this observation, we envision a surrogate model of the deployment platform, that predicts timing uncertainty as a function of global system load (e.g., number of applications and communication activity).
This model would then provide the uncertainty parameters used by the design-time analysis.

In such setting, integrating a new application would first update the estimated platform load, from which timing uncertainty distributions are derived.
The applications can then be analysed individually under these conditions, avoiding full re-validation of the system.
Simple load-to-latency relations (e.g., linear or polynomial models) could serve as a first approximation, while more advanced approaches could learn these relations from monitored execution traces.

\section{Conclusion}
\label{sec:conclusion}

Software-Defined Vehicles (SDVs) require timing analysis methods capable of preserving confidence in system behaviour despite increasing software and middleware complexity.
In this work, we propose a probabilistic timing-analysis framework combining architectural timing models, simulation-based trace generation, and functional-chain reaction-time analysis.
The same analysis pipeline is applied both to simulated executions and to monitored traces collected from a Kuksa-based SDV implementation, enabling direct comparison between design-time predictions and observed runtime behaviour.
The results show that the proposed abstractions remain representative of the implementation behaviour under different middleware load conditions while preserving conservative timing coverage.
Beyond validating the timing abstraction itself, the results suggest that deployment-induced timing uncertainty exhibits sufficiently predictable dependencies on middleware activity to support more incremental timing verification approaches.
Rather than requiring complete system-wide reanalysis after deployment changes, applications may be re-verified individually under updated platform uncertainty conditions.
Future work will focus on extending the evaluation to more complex SDV workloads and on deriving middleware uncertainty models from observed platform activity.

\bibliographystyle{IEEEtran}
\bibliography{ref}

\end{document}